\documentclass[sigconf]{acmart}

\AtBeginDocument{%
  }

\copyrightyear{2024}
\acmYear{2024}
\setcopyright{acmlicensed}\acmConference[CIKM '24]{Proceedings of the 33rd ACM International Conference on Information and Knowledge Management}{October 21--25, 2024}{Boise, ID, USA}
\acmBooktitle{Proceedings of the 33rd ACM International Conference on Information and Knowledge Management (CIKM '24), October 21--25, 2024, Boise, ID, USA}
\acmDOI{10.1145/3627673.3679711}
\acmISBN{979-8-4007-0436-9/24/10}

\acmSubmissionID{1099}

\usepackage{booktabs}
\usepackage{multirow}
\usepackage{diagbox}
\usepackage{color}

\begin{document}

\title{\textbf{Towards Coarse-grained Visual Language Navigation Task Planning Enhanced by Event Knowledge Graph}}

\author{Kaichen Zhao}
\email{22210240394@m.fudan.edu.cn}
\authornote{Equal Contribution.}
\affiliation{
  \institution{Shanghai Key Laboratory of Data \newline Science, School of Computer Science,\newline Fudan University}
  \city{Shanghai}
  \country{China}
}
\author{Yaoxian Song}
\email{songyaoxian@zju.edu.cn}
\authornotemark[1]
\affiliation{
  \institution{Zhejiang University}
  \city{Hangzhou}
  \country{China}
}

\author{Haiquan Zhao}
\email{22210240393@m.fudan.edu.cn}
\affiliation{
  \institution{Shanghai Key Laboratory of Data \newline Science, School of Computer Science,\newline Fudan University}
  \city{Shanghai}
  \country{China}
}

\author{Haoyu Liu}
\email{hyuliu20@gmail.com}
\affiliation{
  \institution{Research Center for Intelligent Robotics, Zhejiang Lab}
  \city{Hangzhou}
  \country{China}
}

\author{Tiefeng Li}
\email{litiefeng@zju.edu.cn}
\affiliation{
  \institution{Zhejiang University}
  \city{Hangzhou}
  \country{China}
}

\author{Zhixu Li}
\email{zhixuli@fudan.edu.cn}
\authornote{Corresponding author.}
\affiliation{
  \institution{Shanghai Key Laboratory of Data \newline Science, School of Computer Science,\newline Fudan University}
  \city{Shanghai}
  \country{China}
}

\begin{abstract}
Visual language navigation (VLN) is one of the important research in embodied AI. It aims to enable an agent to understand the surrounding environment and complete navigation tasks. VLN instructions could be categorized into coarse-grained and fine-grained commands. Fine-grained command describes a whole task with subtasks step-by-step. In contrast, coarse-grained command gives an abstract task description, which more suites human habits. Most existing work focuses on the former kind of instruction in VLN tasks, ignoring the latter abstract instructions belonging to daily life scenarios. To overcome the above challenge in abstract instruction, we attempt to consider coarse-grained instruction in VLN by event knowledge enhancement. Specifically, we first propose a prompt-based framework to extract an event knowledge graph (named \textbf{VLN-EventKG}) for VLN integrally over multiple mainstream benchmark datasets. Through small and large language model collaboration, we realize knowledge-enhanced navigation planning (named \textbf{EventNav}) for VLN tasks with coarse-grained instruction input. Additionally, we design a novel dynamic history backtracking module to correct potential error action planning in real time. Experimental results in various public benchmarks show our knowledge-enhanced method has superiority in coarse-grained-instruction VLN using our proposed VLN-EventKG with over $5\%$ improvement in success rate. Our project is available at \url{https://sites.google.com/view/vln-eventkg}
\end{abstract}



\maketitle

\begin{figure}[h]
    \centering
    \includegraphics[width=0.8\linewidth]{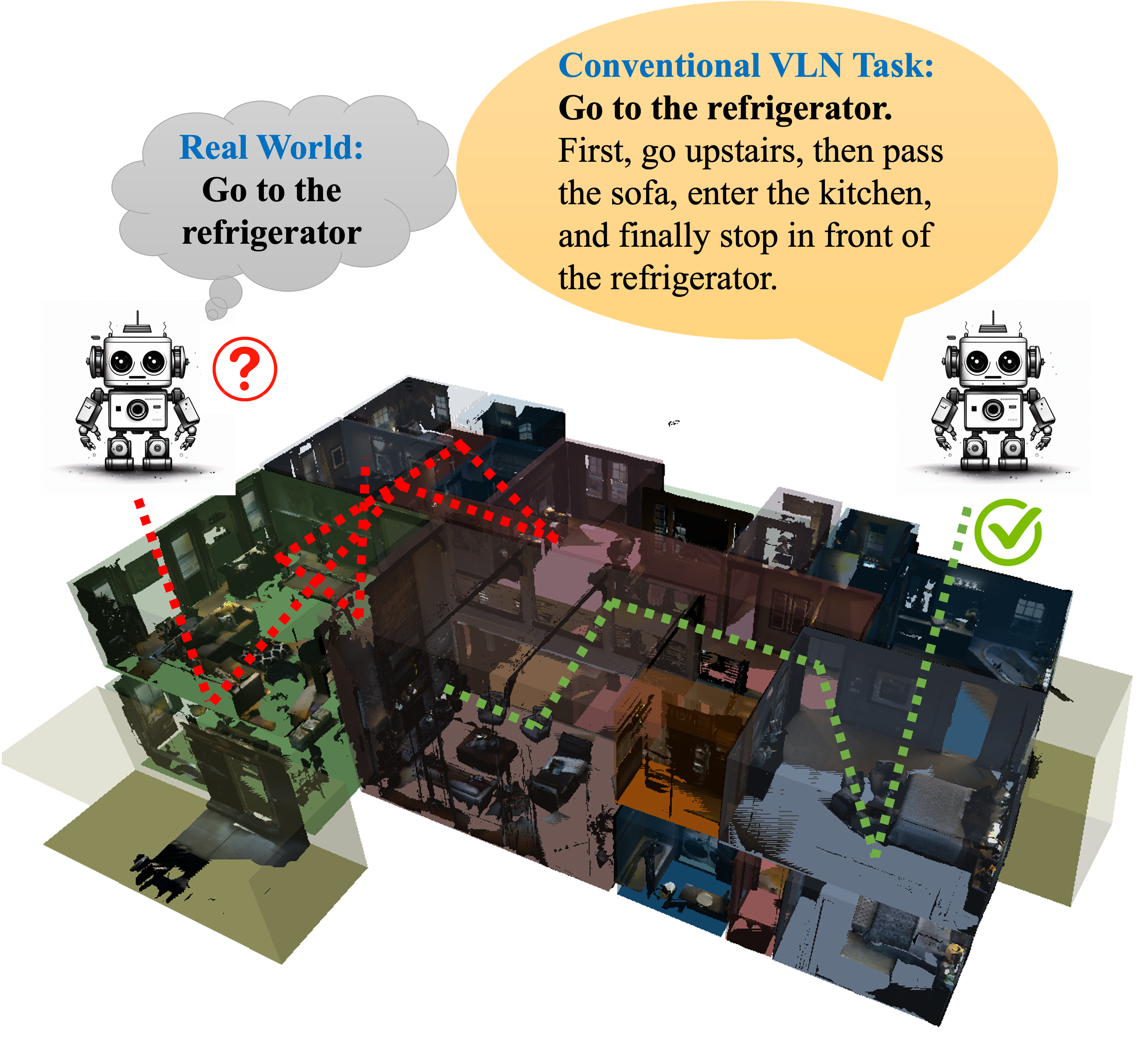}
    \caption{For Conventional VLN tasks, fine-grained task descriptions are provided to the agent. However, in the real world, humans often only provide coarse-grained task descriptions}
    \label{fig:story}
\end{figure}
\section{Introduction}

Visual language navigation (VLN) task~\cite{Seq2Seq-SF,Wang_Huang} aims to enable agents to understand the surrounding environment and accomplish tasks based on human instructions. It is one of the crucial research areas for enabling AI to interact with the real physical world.

\begin{figure*}[t]
    \includegraphics[width=0.9\textwidth]{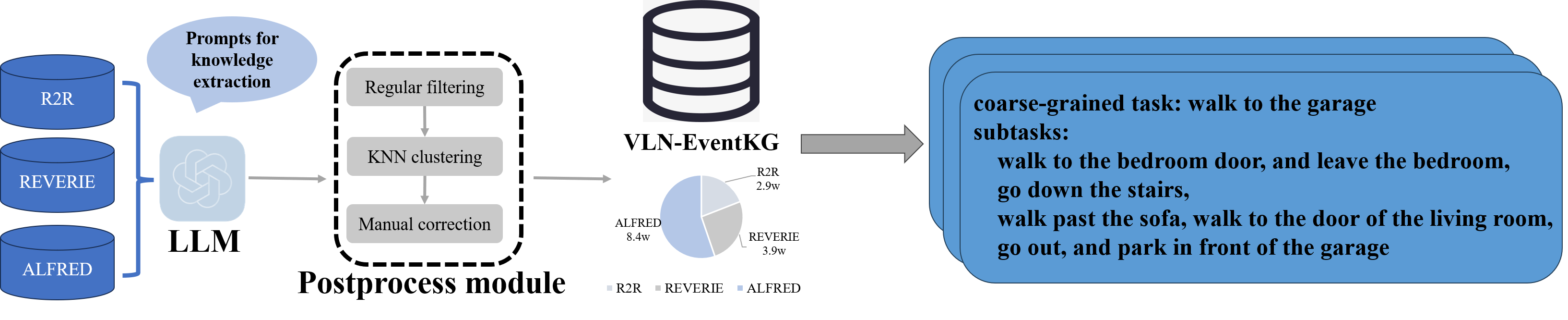}
    \caption{The construction pipeline of the event knowledge graph.}
    \label{fig:knowledge}
\end{figure*}

Existing VLN methods~\cite{Shridhar_Thomason, FILM, HOP, Hong_Wu_Qi_Rodriguez, ET} face the challenges in coarse-grained instructions understanding, in which oracle fine-grained subtask descriptions are supplemented to assist navigation planning. However, for real-world scenarios, instructions in human-robot interaction and navigation are usually abstract and coarse-grained.
For instance, a series of fine-grained detailed instructions make navigation planning easy for an agent ("First, go upstairs, then pass the sofa, enter the kitchen, and finally stop in front of the refrigerator"), however, in real world, humans often give coarse-grained instructions: ("go to the refrigerator"), as shown in Figure.~\ref{fig:story}. The mainstream approaches~\cite{RecBert, ET, FILM} use a transformer architecture model to solve the entire VLN task end-to-end. These models are pre-trained on large VLN datasets using various data augmentation strategies. Extra fine-grained oracle subtask descriptions are used during in training process, which are proven to improve performance significantly compared to models trained without using that~\cite{Shridhar_Thomason}. 

Recently, with the rapid development of large language models, several works~\cite{Micheli_Fleuret, Huang_Abbeel} have explored to use LLM(Large Language Models) for task planning, aiming to exploit coarse-grained instructions to predict fine-grained instructions. However, commonsense knowledge inside the general LLM is not accurately applicable to the specific VLN domain, which also could bring unnecessary noise knowledge leading to bad performance~\cite{Cat-shaped-Mug, zero-shot-planners} on task planning in VLN datasets. Furthermore, most of them focus on the whole performance achievement missing coarse-grained task understanding and task decomposition modeling at the symbolic level.

To make up for the deficiency of abstract and coarse-grained understanding in VLN, we attempt to investigate VLN planning using only coarse-grained task instructions.
Compared to existing VLN research, two key points are considered:
\textbf{1) Coarse-grained instruction understanding}   No oracle fine-grained instructions are used in the modeling stage of VLN planner, which requires the model to understand and reason tasks from abstract and coarse-grained instructions.
\textbf{2) LLM-KG joint VLN model} LLM and structured external knowledge graph are utilized complementarily, which is desired to obtain the reasoning ability of LLM and VLN-domain knowledge from the external knowledge graph eliminating out-of-domain noisy information from general LLM.

For coarse-grained instruction understanding, we propose an event knowledge graph for VLN tasks named \textbf{VLN-EventKG}. Most existing VLN benchmarks are used as data sources to extract event-level knowledge of sequential decisions for VLN planning. Specifically, we perform knowledge extraction on multiple VLN datasets to obtain fine-grained task sequences for each coarse-grained task, in which consequent relations are conceptualized. Our VLN-EventKG is used as an external knowledge graph to guide fine-grained subtasks generation from the input of coarse-grained instructions.
For LLM-KG joint VLN model, we design a large-small-model collaborative framework using event knowledge, named \textbf{EventNav}, in which we extract the event knowledge of VLN scenarios to build a VLN-specific event knowledge graph, used to enhance LLM’s planning capabilities in VLN tasks. In addition, we also propose a dynamic backtracking mechanism to alleviate the accumulation of errors in the model during task execution.

To verify the effectiveness of our proposed method \textbf{VLN-EventKG}, we evaluate our proposed event-knowledge-enhanced model \textbf{EventNav} on public R2R~\cite{R2R}, REVERIE~\cite{FAST-MATTN}, and ALFRED~\cite{Shridhar_Thomason} datasets. Ablation studies about the scale of event knowledge and dynamic backtracking mechanism are conducted to further verify the superiorities of our proposed method.

In summary, the contributions of this paper are as follows:
\begin{enumerate}
    \item We first introduce event knowledge into the sequential decision-making process  (i.e. VLN), consider the correlation of events in the VLN scenario, and guide the model to make correct task planning;
    \item We propose a new event knowledge graph for VLN tasks, which extracts rich event knowledge over public VLN benchmarks, used to help LLM perform more reasonable task planning; 
    \item We propose a novel dynamic backtracking mechanism, allowing the model to assess whether the current task can be successfully executed in real time and backtrack when necessary;
    \item Experimental results on R2R~\cite{R2R}, REVERIE~\cite{FAST-MATTN}, and ALFRED~\cite{Shridhar_Thomason} show that under task settings that only provide coarse-grained task descriptions, our \textbf{EventNav} outperforms the best existing models by more than $5\%$ averagely in success rate.
\end{enumerate}

\section{Related Work}
\textbf{Visual Language Navigation} In recent years, the visual language navigation task has attracted widespread attention from researchers. This task~\cite{Seq2Seq-SF, Wang_Huang} aims to build an intelligent agent that can act in a three-dimensional environment and follow human instructions. Building a system that understands and executes human instructions has been the subject of much previous work. These instruction types include, but are not limited to, structured commands or logic programs~\cite{Manna_Pnueli_1992, Puig_Ra_Boben}, natural language~\cite{Chen_Mooney, Tellex_Kollar}, images~\cite{Lynch_Khansari}, or a mixture of modalities\cite{Lynch_Sermanet}. These efforts focus on mapping the context of instructions and structural words onto final actions. However, in the real world, agents need to be able to process raw sensory input. Therefore, the visual language navigation task introduces rich unstructured visual context to inform the agent's exploration, perception and execution~\cite{Seq2Seq-SF, Chen_Suhr, Krantz_Wijmans}.

\textbf{Conventional Method for VLN} Algorithms in visual language navigation are based on reinforcement learning~\cite{Li_Wang_Tang} or imitation learning~\cite{Speaker-Follwer}. In addition, auxiliary tasks, such as pre-trained on subtasks~\cite{Zhu_Hu_Chen}, speaker-driven route selection~\cite{Speaker-Follwer}, crossmodal matching~\cite{Huang_Jain, RCM}, text-based pre-trained~\cite{Hausknecht_El, Shridhar_Yuan}, progress estimation~\cite{Ma_Lu_Wu, Ma_Wu_AlRegib}, are proposed to improve the performance and generalization of neural agents in seen and unseen environments. For data-centric learning, researchers explore how to utilize data more effectively and synthesize more diverse data. Speaker-follower~\cite{Speaker-Follwer} introduces a speaker model to enhance instruction trajectory pairs. Envdrop~\cite{EnvDrop} breaks the limit of visible environment variability through environment dropout. CCC~\cite{Wang_Liang} aims to learn instruction-following (follower) and instruction generation (speaker) simultaneously. Most of these methods use recurrent neural networks, and the core idea is to encode previous visual observations and behaviors into a hidden state. However, recurrent neural networks have long-range dependency problems and their ability to model long sequences is poor. In recent years, with the introduction of transformer architecture~\cite{Vaswani_Shazeer}, it can provide global-level attention interaction for long sequence tasks. VLN-Bert~\cite{Majumdar_Shrivastava} trains a transformer model to model compatibility between instructions and a set of generated trajectories. Recurrent VLBERT~\cite{Hong_Wu_Qi_Rodriguez} uses explicit recursive state and pre-trained VLBERT to process observations at arbitrary time steps, but this solution is difficult to solve long sequence VLN tasks. Transformers have successfully implemented a wide range of classification and generation tasks, from language to images and videos. In \cite{Parisotto_Song}, the authors show that long-distance task planning using reinforcement learning transformers is challenging and propose a solution. The transformer architecture can also effectively fuse information from different modalities, where a unified transformer model is responsible for solving problems that require multi-modal information, such as visual question answering~\cite{Lu_Batra}, video subtitles and temporal prediction~\cite{Sun_Myers} or retrieval~\cite{Gabeur_Sun}. E.T.~\cite{ET} proposed an algorithm to model the entire VLN task using a unified transformer architecture.

\textbf{LLM-based VLN} In recent years, the birth of LLM has greatly promoted the development of the entire field of natural language processing. Because LLM has rich common sense knowledge and instruction-following capabilities, some work is also actively exploring the application of LLM to visual language navigation. On task. Shah et al.~\cite{Ichter_Levine_Berkeley} adopted GPT-3~\cite{Brown_Mann_Ryder} to identify “landmarks” or sub-objectives, while Huang et al ~\cite{Huang_Mees} focused their efforts on applying LLM to code generation code. Zhou et al~\cite{Zhou_Zheng} used a LLM to extract common-sense knowledge about the relationship between target objects and objects in observations and perform zero-example reasoning object navigation (zson)~\cite{Majumdar_Aggarwal}. These tasks require LLM’s internal knowledge for task planning. The difference from the above work is that we focus on event knowledge specific to VLN scenarios and build it into a knowledge graph in a structured form, thereby assisting LLM in making more accurate mission planning.

\begin{figure*}[t]
    \centering 
    \includegraphics[width=0.8\textwidth]{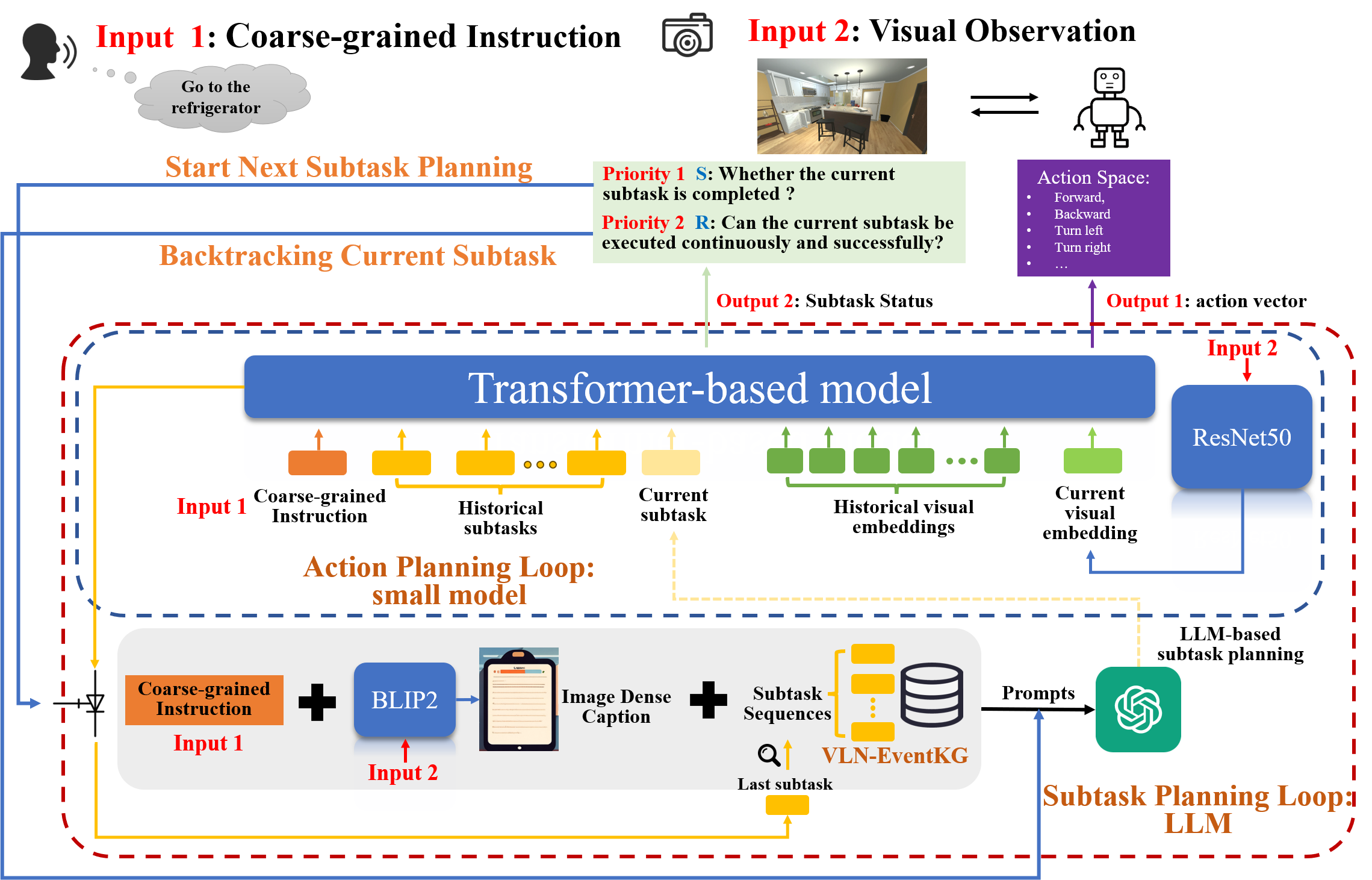}
    \caption{The framework of our proposed EventNav with large-small-model collaboration using event knowledge.}
    \label{fig:model-view}
\end{figure*}

\section{Problem Statement}

The visual language navigation task is designed to allow intelligent agents to interact in the environment according to human instructions to complete corresponding tasks. For each task, it contains a triple $(x_{1:L}, v_{1:t}, a_{1:t})$, consisting of instructions, visual sequences, and action sequences. Among them, the instruction $x_{1:L}$ is a natural language with length $L$. The visual sequence $v_{1:t}$ contains $t$ images, representing the real trajectory during the execution of the task. The action sequences $a_{1:t}$ is a sequence of $t$ action type labels $a_{t} \in \{A\}$ where $A$ is action space. The task goal is to fit a function $f$, whose input is instruction information, visual information, historical actions, and the output is the predicted next action. 

\begin{equation}\label{eq1}
\begin{aligned}
a_t = f(x_{1:L}, v_{1:t}, a_{1:t-1}).
\end{aligned} 
\end{equation}
The above process is carried out iteratively. The agent continuously predicts new actions to obtain new visual environments. It stops until the agent thinks the task is completed or reaches the maximum step limit.

Since $x$ usually contains fine-grained task instructions, and there is a large gap with the real world, we aim to study the VLN algorithm that only contains coarse-grained task instructions. In this task setting, due to the lack of fine-grained instructions, we explore utilizing external event knowledge to enhance navigation planning. Specifically, we construct a novel VLN-specific event knowledge graph and propose an event-enhanced VLN planner to improve VLN performance under the condition of coarse-grained instructions.

\section{Event Knowledge Graph for VLN}
\subsection{Events in VLN Scenery}
An event represents a specific activity or situation that occurs at a specific event and location. Structurally representing events in a knowledge graph, can help the system understand the correlation between events and help model retrieval, reasoning, and analysis. In the VLN scenario, to complete a coarse-grained instruction, the model needs to serially execute a series of fine-grained instructions. We define these fine-grained instructions as event knowledge in VLN tasks. For a mission planning model, only by fully understanding the internal event knowledge of the VLN scenario can better decisions be made. Although the general LLM has rich experimental knowledge, the VLN task has a large amount of prior knowledge that fits the scenery, which is specifically reflected in the correlation between each subtask. This knowledge is not yet available in the general LLM. For example, for the event "pick up an apple", in the real world, subsequent events may include "wash the apple", "eat the apple", "cut the apple", "peel" and so on, but in a specific VLN scenery, the subsequent events may only include "cleaning the apple" and "cutting the apple". This mismatch in event distribution may cause LLM to generate subtasks that do not fit the VLN scenery during prediction. Therefore, event knowledge especially with \textbf{sequential relationships} that fits the VLN scenery can be of great help to LLM in task planning. It can constrain the subtasks predicted by LLM within a small and reasonable range, thereby prompting the LLM to predict more accurate subtasks.

\begin{figure*}[t]
    \includegraphics[width=0.9\textwidth]{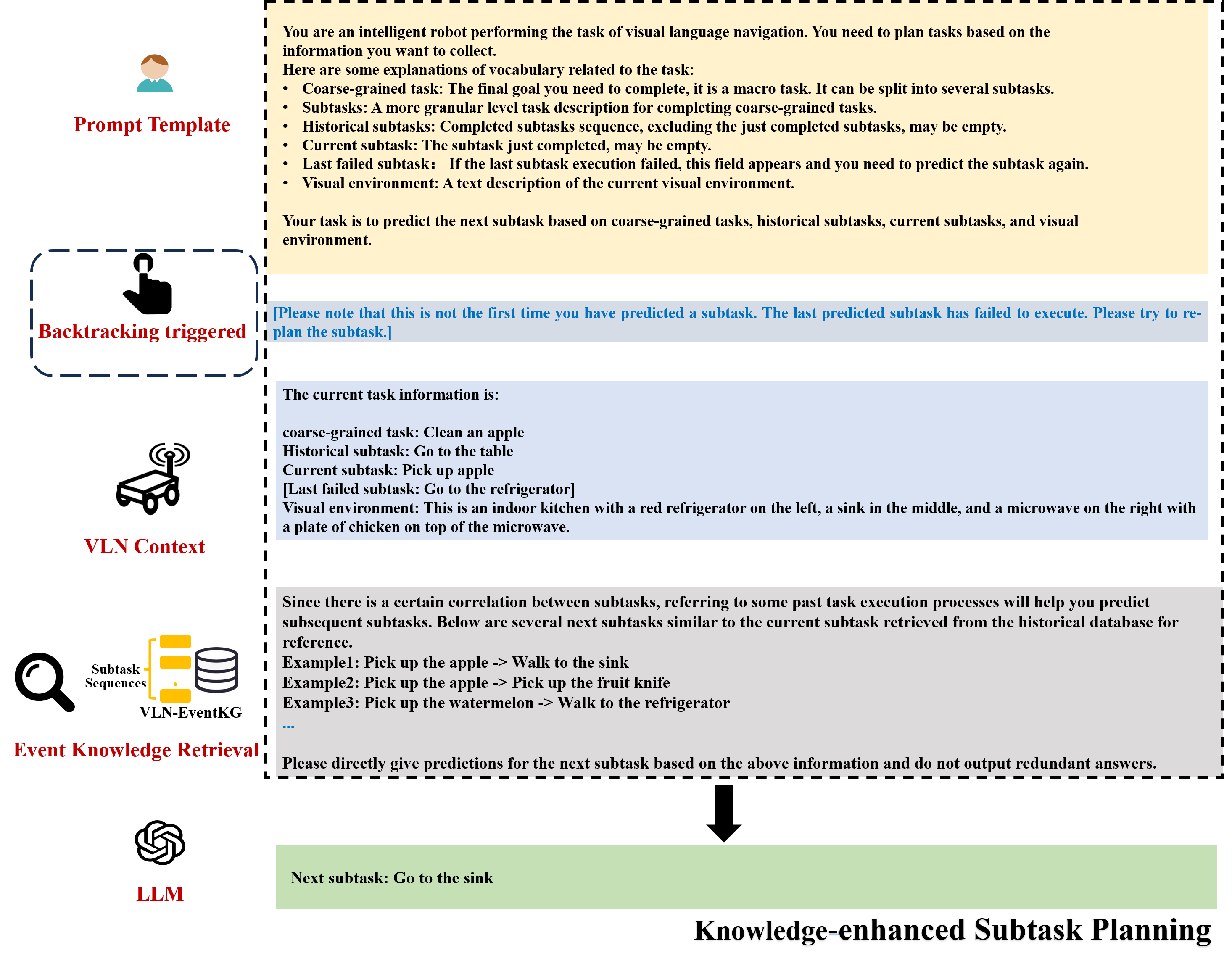}
    \caption{Subtask generation based on event knowledge (VLN-EventKG) retrieval and LLM using promoting learning.}
    \label{fig:prompt2}
\end{figure*}

\subsection{Data Collection and Construction}
In our work, we first integrate event knowledge within current mainstream VLN benchmarks to build a novel VLN-specific event knowledge. Specifically, we try to extract all coarse-grained tasks and corresponding subtask sequences in the dataset. Thus building an event knowledge graph. The event knowledge graph describes the execution sequence of subtasks corresponding to each coarse-grained task. These subtasks serve as external knowledge to assist the task planning of the LLM. We studied three typical VLN datasets: ALFRED~\cite{Shridhar_Thomason}, R2R~\cite{R2R}, and REVERIE~\cite{FAST-MATTN}. In ALFRED~\cite{Shridhar_Thomason}, the corresponding subtask sequence of each coarse-grained task is given in the form of key-value pairs. This is already the result we want, so there is no need for additional extraction work. For the R2R~\cite{R2R} dataset, the coarse-grained tasks and subtasks are a unified text paragraph. For the REVERIE~\cite{FAST-MATTN} dataset, the coarse-grained task and subtask are separated, but the subtask is a text paragraph. Therefore, for the R2R~\cite{R2R} and REVERIE~\cite{FAST-MATTN} datasets, we use LLM to extract coarse-grained tasks and subtasks in them to build an event knowledge graph. Figure ~\ref{fig:knowledge} show the construction process of the event knowledge graph and the prompt design during knowledge extraction respectively. In the end, we get $8.4$w, $2.9$w, and $3.9$w sequential subtasks in the ALFRED~\cite{Shridhar_Thomason}, R2R~\cite{R2R}, and REVERIE~\cite{FAST-MATTN} datasets respectively. By merging the knowledge in the three datasets, we obtained a total of $150$k+ nodes and $120$k+ relationships event knowledge graph. We use the embedding model to embed all nodes into a vector database for subsequent retrieval.

\section{Event-knowledge-enhanced VLN}
\subsection{Framework Overview}
Unlike most algorithms that provide fine-grained task instructions, this work focuses on VLN tasks that only provide coarse-grained instructions. However, simply eliminating all fine-grained subtasks will greatly affect the performance of the entire model. So, we use the powerful task planning capabilities of the LLM to predict subtasks. Besides, we extracted knowledge from the VLN dataset and built an event knowledge graph called VLN-EventKG. We design a collaborative model architecture of large and small models. The architecture consists of two-level loops, the outer is "\textbf{Subtask Planning Loop}" and the inner is "\textbf{Action Planning Loop}". 

For \underline{Subtask Planning Loop}, LLM acts on the outer loop, which obtains three inputs, including the coarse-grained task (\textbf{Input 1}) the image-dense caption after BLIP2 conversion (\textbf{Input 2}), and the similar subtask sequence retrieved from the event knowledge graph \textbf{VLN-EventKG} based on the last subtask. These two inputs with retrieved knowledge are converted into prompts for LLM to predict the next subtask $x_{1:L2}^{current-subtask}$.

For \underline{Action Planning Loop}, according to Eq.~\eqref{eq1}, the small model acts on the inner loop to predict the action (\textbf{Output 1}) $A_t$ at the time $t$ following the input textual instructions $x_{1:L}$, visual observation $v_{1:t}$ and historical actions $a_{1:t-1}$.
Specifically,  \textbf{textual instructions} $x_{1:L}$ can be rewritten into: 
\begin{equation}
\begin{aligned}
x_{1:L} = [x_{1:L1}^{coarse-grained}, x_{1:L2}^{current-subtask}],
\end{aligned} 
\end{equation}
where $x_{1:L1}^{coarsed-grained}$ and  $x_{1:L2}^{current-subtask}$ represent textual instructions. $L1$ and $L2$ indicate the text length of coarse-grained instruction and the current subtask instruction generated by the Subtask Planning Loop. 
\textbf{Visual observation} $v_{1:t}$ includes historical visual information and current visual information at time $t$.
\begin{equation}
\begin{aligned}
v_{1:t} = [v_{1:t-1}, v_{t}],
\end{aligned} 
\end{equation}
Besides, in each step, the small model also outputs two additional signals (\textbf{Output 2}) $S=\{0, 1\}$ and $R \in [0, 1]$. $S$ represents whether the current subtask is completed, and $R$ represents the probability of the current subtask completion. $S$ has higher priority if there is a conflict between $S$ and $R$.


When the small model predicts that the subtask has been completed, $S$ is $1$, it promotes the LLM to plan the next subtask. When the small model predicts a low probability $R$ of the current subtask completion,  the LLM will re-plan the current subtask. We design the above-mentioned process as the dynamic backtracking mechanism in Sec.~\ref{sec:back}.
The overall model architecture is shown in Figure.~\ref{fig:model-view}.

\subsection{Knowledge-enhanced Planner}
The event knowledge graph serves as plug-in knowledge for the LLM. In the task planning process of the LLM, it is to constrain the model to predict subtasks that are more suitable for the VLN scenery. Assuming that the current subtask is $T_i$, the prediction of the next subtask $T_{i+1}$ can be obtained:

\begin{equation}
\begin{aligned}
T_{i+1} &= LLM(D, V, H, L),\\
L &= Retrieval(\text{VLN-EventKG}, T_i),
\end{aligned} 
\end{equation}
where $D$ represents the definition of the coarse-grained task, $V$ represents the image-dense caption after BLIP2~\cite{BLIP2} conversion, and $H$ represents the subtask that has been executed. Figure.~\ref{fig:prompt2} shows the prompt of the LLM used in subtask prediction. Since the subtasks predicted by the LLM may not be correct, the small model will also ask the LLM to re-plan the subtask by triggering the dynamic backtracking mechanism, shown in the blue frame. 

In addition, $Retrieval$ represents the vector retrieval. We use the embedding model to retrieve the current subtask $T_i$ as the query in the Event Knowledge Graph \textbf{VLN-EventKG}. We retrieve $topk$ subtasks $\{s_1, s_2, ..., s_k\}$ similar to the current subtask $T_i$, that is $\{T_{i}^{s_1}, T_{i}^{s_2}, ..., T_{i}^{s_k}\}$, and obtain their subsequent subtasks \\$\{T_{i\rightarrow i+1}^{s_1}, T_{i\rightarrow i+1}^{s_2}, ... ,T_{i\rightarrow i+1}^{s_k}\}$. Ultimately, $L$ is a set of similar subtasks and their subsequent subtasks: 
\begin{equation}
\begin{aligned}
L = \{ (T_{i}^{s_1},T_{i\rightarrow i+1}^{s_1}), (T_{i}^{s_2},T_{i\rightarrow i+1}^{s_2}), ... ,(T_{i}^{s_k},T_{i\rightarrow i+1}^{s_k})\}.
\end{aligned} 
\end{equation}

\subsection{Dynamic History Backtracking Mechanism}\label{sec:back}
Existing visual language navigation methods predict the next action entirely based on language instructions and visual information. It is terminated when the model predicts the task has been completed or execution reaches the maximum step. However, any error during task execution may be continuously amplified and eventually lead to task failure. Under the framework of large and small model collaboration, the LLM is used for subtask planning and predicts the next subtask, while the small model predicts the next action based on coarse-grained tasks, visual information, and subtasks predicted by the LLM. 
Since the subtasks predicted by LLM may not be completely correct, incorrect subtasks may result in incorrect action generation which eventually leads to failure of the navigation task.
However, the subtasks predicted by LLM are not completely fixed. If LLM can adjust the prediction results of the subtasks based on certain signals and guide the small model to perform extra actions, the failure of the task could be avoided. 
Specifically, during the execution of the generated action, in addition to predicting the next action, the small model also predicts two additional values (signals), namely $S$ and $R$. $S$ represents whether the current subtask is completed (completed, set $1$, and vice versa), and $R$ represents the probability of the current subtask completion.

\begin{figure}[t]
    \centering
    \includegraphics[width=0.8\linewidth]{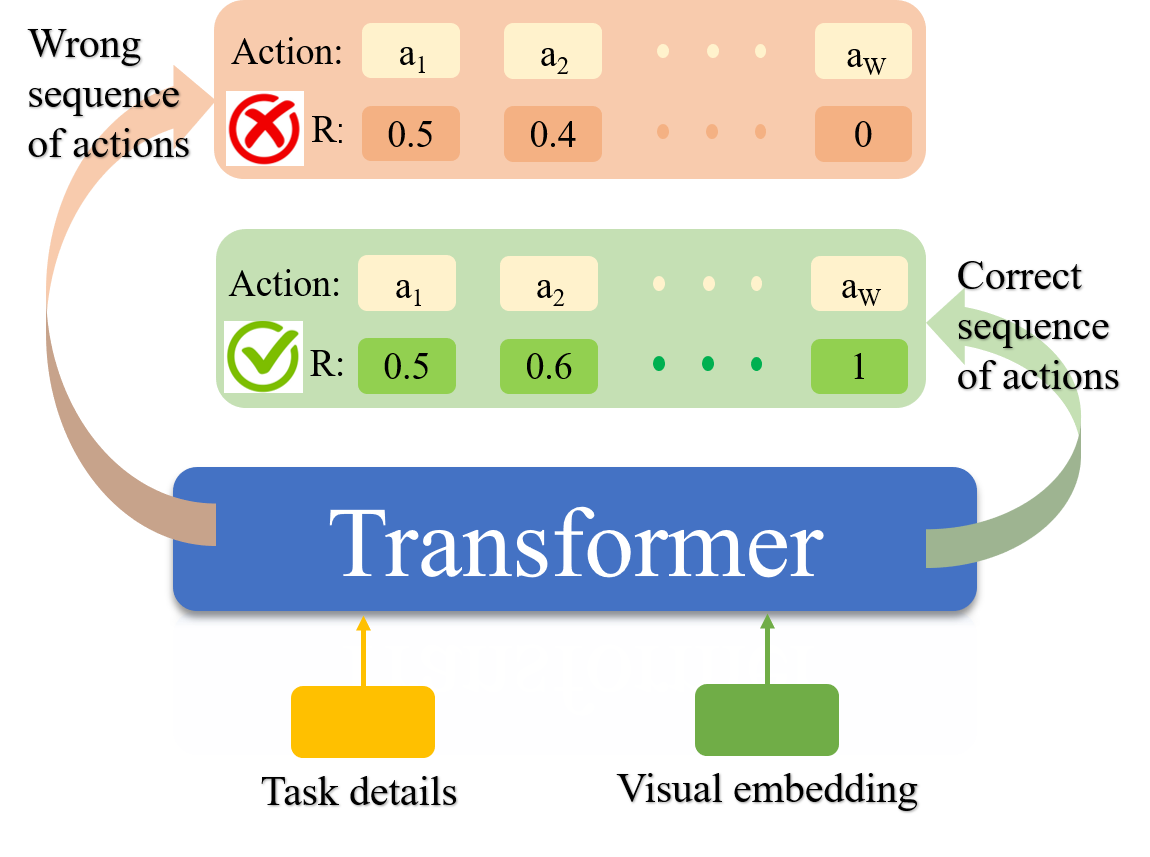}
    \caption{The training details of the dynamic backtracking mechanism.}
    \label{fig:backtrack}
\end{figure}

The prediction with probability $R$ is the triggering condition for the backtracking mechanism. During the model training process, we obtain negative samples by random sampling from action space.
Specifically, we set that the subtask requires $W$ actions to complete and the start point value of $R_1 = 0.5$.
For the positive action trajectory, after the $i-th$ action of the subtask is completed, $R_i = R_1 + \frac{i}{2W}, i \leq W$. For negative action trajectory, we use weak negative samples to approximate real negative samples, by random sampling within action space, then the corresponding $R_i = R_{0}-\frac{i}{2W}, i \leq W$. The whole process is shown in  Figure.~\ref{fig:backtrack}.


\begin{table}[t]
    \centering
    \begin{tabular}{cc}
        \toprule
        Dataset  & $D_{avg}$ \\
        \midrule
        R2R~\cite{R2R}  & 2.35 \\
        REVERIE~\cite{FAST-MATTN}  & 3.57\\
        ALFRED~\cite{Shridhar_Thomason}  & 8.26 \\
        \bottomrule
    \end{tabular}
\caption{The average number of actions required for each subtask in each dataset benchmarks}
\label{tab:subtask avg}
\end{table}

During model inference, two priority conditions are set to promote a new subtask.

\textbf{Priority 1:}  $S$ predicted to be $1$. This condition has higher priority compared to another one, which means the current subtask has been completed and the LLM can directly predict the next subtask based on the current completed subtask context. We use a transistor switch flag to visualize this process, shown on the left of Figure. \ref{fig:model-view}.

\textbf{Priority 2:} $R$ shows a continuous downward trend or the value of $R$ is less than $0.25$. It means the current subtask could not be completed, in which the agent needs to go back to the start position of the current subtask, and the LLM is prompted to re-plan the subtask based on the last effective subtask context. A specific prompting case is shown in Figure.~\ref{fig:prompt2}.







\begin{table*}[h]
\centering
\resizebox{0.85\textwidth}{!}{
\begin{tabular}{c|cccc|cccc|cccc}
    \hline
 \multirow{2}{*}{Methods}
     & \multicolumn{4}{c|}{Validation Seen} 
     & \multicolumn{4}{c|}{Validation Unseen} 
     & \multicolumn{4}{c}{Test Unseen} \\ 
    & SR$\uparrow$ & NE$\downarrow$ & SPL $\uparrow$ & TL & SR$\uparrow$ & NE$\downarrow$ & SPL $\uparrow$ & TL & SR$\uparrow$ & NE$\downarrow$ & SPL $\uparrow$ & TL \\ 
\hline
     Random & 16 & 9.45 & - & 9.58 & 16 & 9.23 & - & 9.77 & 13 & 9.79 & 12 & 9.89 \\ 
     Human & - & - & - & - & - & - & - & - &  86 & 1.61 & 76 & 11.85 \\ 
\hline
     Seq2Seq-SF\cite{Seq2Seq-SF}& 39 & 6.01 & 40 & 11.33 & 22 & 7.81 & 19 & 11.67 & 20 & 7.85 & 18 & 8.13\\
     Speaker-Follower\cite{Speaker-Follwer}& 54 & 4.29 & 49 & 12.58 & 29 & 7.92 & 27 & 13.02 & 28 & 7.84 & 24 & 15.19 \\
     EnvDrop\cite{EnvDrop}& 55 & 3.82 & 56 & 9.69 & 39 & 5.92 & 36 & 10.24 & 41 & 6.02 & 43 & 11.24\\
     RecBERT\cite{RecBert}& 67 & 3.49 & 63 & 11.45 & \underline{58} & 4.54 & \textbf{54} & 12.94 & \underline{58} & 4.61 & 53 & 12.46\\
     HOP\cite{HOP}& 68 & 3.23 & \textbf{65} & 11.37 & 57 & 4.52& 50 & 12.79 & 56 & 4.39 & \textbf{56} & 12.93\\
     AirBERT\cite{AirBert}& \underline{69} & \underline{3.19} & \underline{64} & 11.56 & 55 & \underline{4.29} & \underline{51}& 12.56 & 55 & \underline{4.33} & \underline{52} & 12.57\\
 \hline
     \textbf{EventNav (ours)}& \textbf{72} & \textbf{2.77} & 63 & 12.58 & \textbf{60} & \textbf{4.23} & \underline{51} & 14.56 & \textbf{59} & \textbf{4.25} & 49 & 14.55 \\
 \hline
\end{tabular}}
\caption{Comparative results between our method and other mainstream methods on the R2R~\cite{R2R} dataset that only provides coarse-grained instructions.}
\label{table:R2R result}
\end{table*}

\begin{table*}[h]
\centering
\resizebox{0.85\textwidth}{!}{
\begin{tabular}{c|cccc|cccc|cccc}
    \hline
 \multirow{2}{*}{Methods}
     & \multicolumn{4}{c|}{Validation Seen} 
     & \multicolumn{4}{c|}{Validation Unseen} 
     & \multicolumn{4}{c}{Test Unseen} \\ 
    & SR$\uparrow$ & OSR$\uparrow$ & SPL$\uparrow$ & TL & SR$\uparrow$ & OSR$\uparrow$ & SPL$\uparrow$ & TL & SR$\uparrow$ & OSR$\uparrow$ & SPL$\uparrow$ & TL \\ 
\hline
    Random & 3 & 8.92 & 2& 11.99 & 2 & 11.93 & 1 & 10.76 & 2 & 8.88 & 1 & 10.34\\
    Human & - & - & - & - & - & - & - & - & 81& 86.83 & 54& 21.18 \\ 
\hline
    RCM\cite{RCM}& 23& 29.44 & 22& 10.70 & 9& 14.23 & 7& 11.98 & 8& 11.68 & 7& 10.60  \\
    FAST-MATTN\cite{FAST-MATTN}& 50& 55.17 & 26& 16.35 & 14& 28.20 & 6& 29.70 & 14& 23.36 & 9& 30.69 \\
    AirBert\cite{AirBert}& 47& 48.98 & 42& 15.16 & 28& 34.51 & 22& 18.71 & \underline{30}& \underline{34.20} & 23& 17.91\\
    RecBERT\cite{RecBert}& 51& 53.90 & \underline{48}& 13.44 & 31& 35.02 & \underline{25}& 16.78 & 30& 32.91 & \textbf{24}& 15.86\\ 
    HOP\cite{HOP}& 53& 54.88 & 37& 13.80 & \underline{32}& \underline{36.24} & \textbf{26}& 16.46 & 30& 33.06 & \textbf{24}& 16.38 \\
    ORIST\cite{ORIST} & 45& 49.12 & 42& 10.73 & 17& 25.02 & 15& 10.90 & 22& 29.20 & 19& 11.38 \\
    CKR\cite{CKR} & \underline{57}& \underline{61.91} & \textbf{53}& 12.16 & 19& 31.44 & 11& 26.26 & 22& 30.40 & 14& 22.46 \\
 \hline
     \textbf{EventNav (ours)}& \textbf{61}& \textbf{63.29} & 45& 15.25 & \textbf{34}& \textbf{39.28} & 23& 16.25 & \textbf{34}& \textbf{39.36} & 23& 16.10 \\
 \hline
\end{tabular}}
\caption{Comparative results between our method and other mainstream methods on the REVERIE~\cite{FAST-MATTN} dataset that only provides coarse-grained instructions.}
\label{table:REVERIE result}
\end{table*}

\begin{table*}[h]
\centering
\resizebox{0.7\linewidth}{!}{
\begin{tabular}{c|cccc|cccc}
    \hline
 \multirow{2}{*}{Methods}
     & \multicolumn{4}{c|}{Test Seen} 
     & \multicolumn{4}{c}{Test Unseen} \\

    & SR$\uparrow$ & GC$\uparrow$ & PLWSR$\uparrow$ & PLWGC$\uparrow$ & SR$\uparrow$ & GC$\uparrow$ & PLWSR$\uparrow$ & PLWGC$\uparrow$ \\
\hline
     Human & - & - & - & - & 91 & 94.5 & - & -\\
\hline
     SEQ2SEQ\cite{Shridhar_Thomason} & 3& 8.00 & 7.29 & 12.56 & 1& 7.30 & 2.42 & 9.09\\
     E.T.\cite{ET} & 22& 29.31 & \textbf{12.39} & 17.64 & 4& 8.60 & 5.77 & 10.24 \\ 
     MOCA\cite{MOCA} & 22& 28.37 & \underline{12.36} & \textbf{26.78} & 5& \underline{14.30} & 9.24 & \underline{19.19}\\
     FILM\cite{FILM} & \underline{26}& 36.15 & 10.39 & 19.17 & - & - & - & -\\  
     EmBert\cite{EmBert} & 25 & \underline{37.69} & 11.29 & \underline{25.60} & \underline{7} & 12.49 & \textbf{9.79} & 18.22 \\
\hline
    \textbf{EventNav (ours)} & \textbf{31}& \textbf{42.20} & 11.97 & 24.62 & \textbf{10}& \textbf{18.75} & \underline{9.26} & \textbf{19.48} \\  
\hline
\end{tabular}}
\caption{Comparative results between our method and other mainstream methods on the ALFRED~\cite{Shridhar_Thomason} dataset that only provides coarse-grained instructions.}
\label{table: ALFRED result}
\end{table*}

\section{Experiment and Analysis}
\subsection{Datasets and Simulation Environments}
All VLN datasets need to be built based on a certain simulation environment. The R2R~\cite{R2R} and REVERIE~\cite{FAST-MATTN} datasets are built on MatterPort3D simulation. In these datasets, the agent can navigate the house according to the navigation graph. The navigation graph consists of viewpoints and edges. Each viewpoint contains a panoramic view, and there is a bidirectional edge between any two navigable viewpoints, that is, the agent can move bidirectionally between two adjacent viewpoints. Initially, the agent is placed at a random viewpoint and inputs a natural language instruction. At each step, $t$, the agent obtains a panoramic graph $O_t = \{o_t, i\}, 1 \leq i \leq 36$, which includes several local views. Each local view $o_t$ represents a navigable viewpoint, and the agent needs to choose one of the viewpoints to go.

The ALFRED~\cite{Shridhar_Thomason} dataset is built on AI2THOR simulation. In this dataset, the agent is placed at a random position, and in each step, a defined action is selected for execution. These actions include two types: movement and navigation. There are $6$ types of movement actions (i.e. forward, backward, left turn, right turn, head up, head down), and $7$ types of interactive actions (i.e. picking up, putting down, opening, closing, washing, cooking, and chopping). The $7$ types of interactive actions involve interacting with target objects in the environment. The target object of the action is represented by a mask. Therefore, for interactive actions, the agent needs to output not only the action category but also the mask of the interactive object.

\subsection{Implementation Details}\label{sec:details}
The framework of our proposed method includes "Subtask Planning Loop" and "Action Planning Loop". LLM (i.e. ChatGPT~\cite{Brown_Mann_Ryder}) is used in "Subtask Planning Loop". We follow the work of~\cite{zhou2024navgpt} and use the BLIP2\cite{BLIP2} to provide LLM with an image-dense caption of the current visual environment. For the event knowledge graph VLN-EventKG, we use the bge-large-en~\cite{BGE} semantic model to perform vector similarity retrieval and retrieve similar subtasks to prompt the LLM for task planning.

In "Action Planning Loop", we follow the work of~\cite{ET} and adopt a unified transformer-based model. For all visual images, we first feed them into ResNet50~\cite{resnet} to obtain the visual representation vector, and then the vector is used as a token for the transformer-based model. In addition, for the ALFRED~\cite{Shridhar_Thomason} dataset benchmark, its tasks not only require the agent to move but also require it to interact with the environment. Therefore, the model also needs to generate a mask to represent which part to interact with. We use ResNet-50 Mask R-CNN~\cite{maskrcnn} to generate a target mask. Overall, only the transformer-based model and Mask R-CNN~\cite{maskrcnn} are involved in the training process, and the rest of the parameters of the models are frozen. We train our proposed model for $20$ epochs using Pytorch $1.13$ on four V100 GPU platforms, with batch size $64$, Adam optimizer, and learning rate $1e-3$.

During the inference process, the LLM is used to subtask planning, and the small model is used to generate actions to be executed. For knowledge enhancement, we retrieve $topk$ similar subtasks in VLN-EventKG each time. Here, we choose $topk = 5$. In the dynamic backtracking mechanism, when the probability $R$ predicted by the small model meets one of the following two conditions, the LLM will be required to re-plan the subtask: 1) $R$ < $0.25$; 2) $R$ shows a downward trend $W$ consecutive times. For $W$ in Figure~\ref{tab:subtask avg} and Sec.~\ref{sec:back}, The average execution length $D_{avg}$ of each subtask over different datasets is shown in Table~\ref{tab:subtask avg}. We set $W$ with weighted $D_{avg}$ for different datasets (i.e.  R2R~\cite{R2R}: $2\times D_{avg}$, REVERIE~\cite{FAST-MATTN} and ALFRED~\cite{Shridhar_Thomason}: $D_{avg}$).

\subsection{Evaluation Metrics}
For R2R~\cite{R2R}:
\begin{itemize}
\item Trajectory Length (TL): the average path length in meters;
\item Navigation Error (NE): the average distance between the agent's final position and the target in meters;
\item Success Rate (SR): the ratio of stopping within $3$ meters to the target; 
\item Success Path Length (SPL): the success rate weighted by the normalized inverse of the Path Length. 
\end{itemize}

\noindent For REVERIE~\cite{FAST-MATTN} also employs Oracle Success Rate (OSR): the ratio of having a viewpoint along the trajectory where the target object is visible.

\noindent For ALFRED~\cite{Shridhar_Thomason}:
\begin{itemize}
\item Success Rate (SR): the success rate of the total task; 
\item Conditional Success (GC): the ratio of subtasks completed over the whole task;
\item Path length weighted SR (PLWSR): the ratio of (length of the expert path) and (length taken by the agent);
\item Path length weighted GC (PLWGC): the ratio of (length of the ground truth Path) and (length taken by the agent). 
\end{itemize}

\subsection{Comparisons on Different VLN benchmarks}
\begin{figure}[t]
    \includegraphics[width=\linewidth]{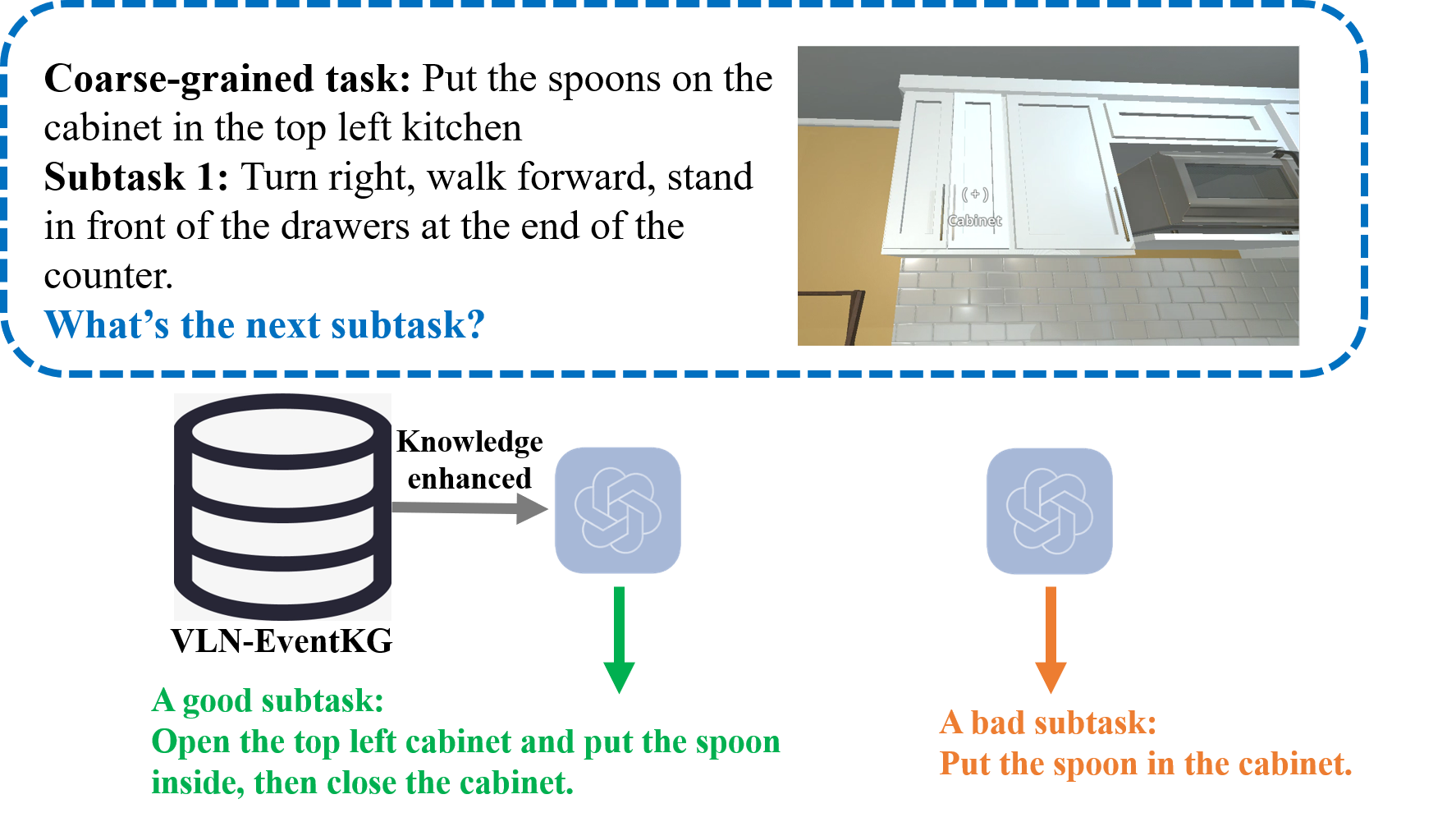}
    \caption{Task planning case study under knowledge enhancement}
    \label{fig:case_study}
\end{figure}

Based on the setting of coarse-grained task descriptions, We evaluate the performance of \textbf{EventNav} and mainstream VLN models on three dataset benchmarks: R2R~\cite{R2R}, REVERIE~\cite{FAST-MATTN} and ALFRED~\cite{Shridhar_Thomason}. Table~\ref{table:R2R result} presents the comparative results between our method and other VLN models in R2R dataset~\cite{R2R} and our proposed model achieves strong performance in most metrics. The scores of SR show an average improvement of \textbf{2\%} over the existing results. In Table~\ref{table:REVERIE result}, the most significant improvement is observed in the unseen test dataset of REVERIE~\cite{FAST-MATTN}, with a \textbf{4\%} increase in SR metric. Compared to the existing methods on SR and OSR metrics, our model achieves an average improvement of \textbf{3.3\%} and \textbf{3.2\%} improvements, respectively. The result of ALFRED~\cite{Shridhar_Thomason} is shown in Table~\ref{table: ALFRED result}. We evaluate performance in two settings with four metrics and achieve SOTA in five metrics. Compared with the existing models, the SR of our model is improved by \textbf{5\%} and \textbf{3\%} on Test Seen and Test unseen settings. From the results shown in the above three tables on different dataset benchmarks, we can find that our VLN-EventKG provides VLN planners with useful event knowledge, and coupled with the dynamic backtracking mechanism, our proposed \textbf{EventNav} model for VLN task achieve competitive performance with coarse-grained instructions input. 
Figure~\ref{fig:case_study} also shows a case study on the coarse-grained VLN task, with the enhancement of the event knowledge graph of VLN-EventKG, LLM can obtain the event knowledge of the VLN scenario, thereby realizing a more reasonable task planning.

\subsection{Ablation Study}

\textbf{VLN-EventKG} For the utilization of event knowledge, we consider the impact of the event knowledge graph VLN-EventKG on the performance of the model. We use a small model that only provides coarse-grained instructions as a baseline (short as \textbf{base}) to consider the following situations:

\begin{enumerate}
    \item \textbf{base+planD}: Without using the event knowledge graph, let LLM plan subtasks \textbf{D}irectly;
    \item \textbf{base+planS}: Knowledge is extracted \textbf{S}eparately for each dataset benchmark, and the event knowledge graph corresponding to each dataset benchmark is used during inference;
    \item \textbf{base+planF}: \textbf{F}using the event knowledge from VLN dataset benchmark domains and using the entire event knowledge graph during inference.
\end{enumerate}


As shown results of Success Rate (SR) in Table~\ref{table:Knowledge base}, we find that in R2R~\cite{R2R}, REVERIE~\cite{FAST-MATTN}, and ALFRED~\cite{Shridhar_Thomason} benchmark, the models fused event knowledge graph (\textbf{base+planF}) perform better than the single event knowledge graph (\textbf{base+planS}). In particular, for R2R~\cite{R2R} and REVERIE's~\cite{FAST-MATTN}, the model fused event knowledge graph has significantly improved the model effect, with the average SR improvement over $2\%$. We attribute it to the similar distributions of task descriptions in above mentioned two datasets. Therefore, the models fused event knowledge graph can obtain broader and diverse external knowledge promoting the next subtask from the LLM. In contrast, subtasks in the ALFRED~\cite{Shridhar_Thomason} dataset benchmark mainly involve interacting with the environment, which exists relative differences from the first two benchmark domains. This reduces the effectiveness of event knowledge to improve performance in the ALFRED~\cite{Shridhar_Thomason} benchmark.

When not using the event knowledge graph and letting LLM plan subtasks directly (base+planD), we found that due to the lack of event knowledge from the knowledge graph, the next subtask predicted by the model is usually very different from the original subtask distribution of the dataset benchmark. This approach could damage the performance of the action prediction for the small model. 
Therefore, our VLN-EventKG plays a constraint role in the subtask planning process based on coarse-grained instructions, which guarantees the LLM plan subtasks close to the distribution of the original dataset benchmark, thereby promoting the abstract task understanding and action generation of the small model.

\textbf{Dynamic Backtracking Mechanism} Dynamic backtracking mechanism aims to determine whether the current subtask can be executed successfully based on the probability level $R$ predicted by the small model, and guide the LLM to re-plan the subtask at the appropriate time. Specifically, when one of the following two conditions occurs, the backtracking mechanism is triggered, denoted in Sec.~\ref{sec:details}: 
\begin{enumerate}
    \item $R < x$;
    \item $R$ shows a downward trend $W$ consecutive times.
\end{enumerate}
where $x$ and $W$ are two hyperparameters of the model inference process. The triggered timing of backtracking has a great impact on the overall performance of the model. Backtracking too early could cause the model to exit early on the correct trajectory while backtracking too late could fail because the maximum step size specified by the task is reached. 

For different dataset benchmarks, we choose different values of $x$ and $W$ to conduct ablation studies. We select the values of $x$ as $(0.1, 0.25, 0.5)$ respectively. We choose the values of $W$ as $(0.5\times D_{avg}, D_{avg}, 2\times D_{avg}, 4\times D_{avg})$, where $D_{avg}$ denotes the average length of its subtasks for a specific dataset in Table~\ref{tab:subtask avg}.

Table~\ref{table:Backtracing} shows the Success Rate (SR) corresponding to different $x$ and $W$ under different dataset benchmarks.
The value of $0.25$ for $x$ gave the best model results. In addition, the best results are obtained when $W$ is $2\times D_{avg}$ for R2R~\cite{R2R} and $D_{avg}$ for REVERIE~\cite{FAST-MATTN} and ALFRED~\cite{Shridhar_Thomason}. 
Twice $D_{avg}$ used in R2R~\cite{R2R}, we think it is because the lengths of subtask steps in this benchmark are shorter than the other two, and selecting $W = D_{avg}$ could cause the model to early backtrack to terminate the correct action planning loop. 
In contrast, $W$ should be appropriately selected for a larger value for REVERIE~\cite{FAST-MATTN} and ALFRED~\cite{Shridhar_Thomason} benchmarks due to the longer average subtask steps.
When $R$ decreases $D_{avg}$ times continuously can roughly determine that a subtask will eventually fail, so $W$ should be $D_{avg}$. 
In addition, Table~\ref{table:Knowledge base} also shows the impact of the dynamic backtracking mechanism on the overall success rate of the model. Our method ($base+planF+backtrace$) integrates the dynamic backtrace mechanism with the whole event knowledge graph showing significant improvement in the success rate of over all VLN benchmarks.

\begin{table}
    \centering
    \scalebox{0.85}{

    \begin{tabular}{cccc}
    \hline
        Method/Dataset & R2R~\cite{R2R} & REVERIE~\cite{FAST-MATTN} & ALFRED~\cite{Shridhar_Thomason} \\
    \hline
        \textbf{base} & 55& 22& 14\\
        \textbf{base+planD} & 49& 20& 13\\
        \textbf{base+planS} & 61& 45& 26\\
        \textbf{base+planF} & 63& 48& 26\\
        \textbf{base+planF+backtrace (Ours)} & \textbf{72} & \textbf{61} & \textbf{31}\\
    \hline
    \end{tabular}}
\caption{The impact of event knowledge graph on model task planning effect.}
\label{table:Knowledge base}
\end{table}

\begin{table}[h]
\centering
\begin{tabular}{c|ccc|ccc|ccc}
    \hline
 \multirow{2}{*}{\diagbox{$W$}{$x$}}
     & \multicolumn{3}{c|}{R2R~\cite{R2R}} 
     & \multicolumn{3}{c|}{REVERIE~\cite{FAST-MATTN}} 
     & \multicolumn{3}{c}{ALFRED~\cite{Shridhar_Thomason}} \\
    & 0.1 & 0.25 & 0.5 & 0.1 & 0.25 & 0.5 & 0.1 & 0.25 & 0.5  \\
\hline
     0.5$\times D_{avg}$ & 63 & 66 & 60 &  52&  58&  55&  27&  29&  22\\
     $D_{avg}$ & 69 & 69 & 67 &  59&  \textbf{61}&  51&  30&  \textbf{31}&  29\\
     2$\times D_{avg}$ & 71 & \textbf{72} & 70 &  58&  60&  51&  26&  30&  25\\
     4$\times D_{avg}$ & 70 & 70 & 69 &  58&  55&  50&  24&  27& 25\\
 \hline
\end{tabular}
\caption{In the dynamic backtracking mechanism, the impact of $x$ and $W$ on the experimental results.}
\label{table:Backtracing} 
\end{table}

\section{Conclusion}
In this paper, we investigate coarse-grained VLN planning guided by event knowledge. Event knowledge especially consequent relations is considered for sequential decisions. Specifically, we propose the first event knowledge graph \textbf{VLN-EventKG} for VLN tasks by extracting and conceptualizing the activity sequences across various public VLN benchmarks. An event-knowledge-enhanced VLN planning model is designed under the large-small-model collaborative framework \textbf{EventNav} to realize coarse-grained instruction understanding and decomposition. A dynamic backtracking mechanism is considered to further improve the success rate of VLN by in-time correction of intermediate decisions. Experimental results indicate the importance of event knowledge in sequential decisions, (i.e. VLN planning). Our proposed \textbf{VLN-EventKG} combined with \textbf{EventNav} effectively improve abstract human instructions understanding and hierarchical task planning by over $5\%$ average success rate in various public benchmarks.

\section{Acknowledgments}
We thank the anonymous reviewers for their valuable comments. This work is supported by Postdoctoral Fellowship Program of CPSF under Grant Number GZC20232292, National Natural Science Foundation of China (No.62072323, U21A20488, No.62102276), Shanghai Science and Technology Innovation Action Plan (No.22511104700), China Postdoctoral Science Foundation (Grant No. 2023M732563), and Zhejiang Lab Open Research Project (No.K2022NB0AB04).


\end{document}